\documentclass[12pt]{article}
\usepackage{graphicx} 
\usepackage{amsmath, amssymb, amsthm}
\usepackage{hyperref}
\usepackage{authblk}
\usepackage[dvipsnames]{xcolor}
\usepackage{indentfirst}
\usepackage{float}
\usepackage{breakcites}
\usepackage{titlesec}
\usepackage{changepage}
\usepackage[square,numbers]{natbib}
\hypersetup{
    colorlinks=true,
    linkcolor=black,
    filecolor=blue,      
    urlcolor=blue,
    citecolor=blue,
    }

\title{\huge\textbf{A Comprehensive Analysis of the Future of Atomically Precise Manufacturing}}

\author{\large Vadym Shvydun$^1$, Gabriel Bristot$^2$, and Justin Sato$^3$ \\ \footnotesize $^1$vadim.shvydun@gmail.com \hspace{0.1cm} $^2$gbristot2000@gmail.com \hspace{0.1cm} $^3$justinsccs1@gmail.com}

\date{August 2024}

\titleclass{\subsubsubsection}{straight}[\subsection]

\titleformat{\subsubsubsection}
  {\normalfont\normalsize\bfseries}{\thesubsubsubsection}{1em}{}
\titlespacing{\subsubsubsection}
{0pt}{3.25ex plus 1ex minus .2ex}{1.5ex plus .2ex}

\makeatletter
\renewcommand\paragraph{\startsection{paragraph}{4}{\z@}%
            {-2.5ex\plus -1ex \minus -.25ex}%
            {1.25ex \plus .25ex}%
            {\normalfont\normalsize\bfseries}}
\makeatother
\begin{document}

\newcommand{\citetb}[1]{\textcolor{blue}{\citet{#1}}}

\pagenumbering{arabic}

\maketitle

\begin{adjustwidth}{1.5cm}{1.5cm} 
\tableofcontents
\end{adjustwidth}

\newpage

\section{Abstract}

Atomically Precise Manufacturing (APM) refers to the assembly of materials with atomic precision, representing a highly advanced technology with significant potential. However, the development of APM remains in its early stages, with applications largely confined to specialized fields and lacking cohesion within a unified discipline. The current literature on APM is often dominated by older, speculative papers that discuss its immense potential risks and benefits without sufficient grounding in the latest advancements or practical limitations that exist today. This paper aims to bridge this gap by providing a comprehensive assessment of current APM and near-APM technologies, as well as using the barriers to further progress to predict future developments. Through this analysis, we seek to establish a clearer understanding of the present state of the technology and then use these insights to predict the future trajectory of APM. By doing so, we aim to create a more grounded discourse on APM and its potential risks and benefits, while also guiding future research on the necessary regulations and safety considerations for this emerging field.

\section{Introduction}

Currently, APM-specific literature considers a singular speculative technology representing the most futuristic or extreme scenarios, such as (\citetb{Umbrello}), attempting to predict net societal impact of APM, or (\citetb{Jacobstein2010ForesightGF}), providing guidelines for speculative future development. While these papers provide an intriguing perspective of the potential that APM has, they also tend to skew the discourse towards very idealized, impractical outcomes, some of which may not be possible in the near future due to inherent physical constraints. Furthermore, many of these papers were written before the development of many APM technologies that exist today, which have since given a better understanding of APM’s future. 

This focus on a singular type of speculative APM has led to many problems. First, the literature often talks about the most extreme possibilities, such as fully autonomous self-assembling nanobots or doomsday scenarios (\citetb{drexlerconvo}) without considering the intermediate technologies or incremental steps to get to the point of those extreme possibilities. A justification for considering these extreme possibilities is that if they are not addressed now (while still largely speculative), by the time the technologies have arrived it is too late. Even a small chance of an extinction event has massive counterfactual value. However, this speculation work has led to minimal impact on the modern day development scene of APM, especially because it has gone in a completely different direction compared to the speculative extreme one described in past research. 

This gap between theory and practice is a problem in the field of APM. While theoretical work has often focused on the future of APM—such as autonomous self-replicating nanobots, mass surveillance, or boundless material wealth (\citetb{Umbrello})—there is a stark disconnect between these speculative visions and the current state of practical applications. Much of the existing literature fails to offer concrete guidance for developing APM in ways that mitigate these extreme risks, likely due to the uncertainty and speculative nature of the field at the time those papers were written. In essence, this speculative approach chooses to ignore the bridge from now to the future of the technology, leading to a lack of any actionable steps or even a clear outline on how to change courses on the development of APM.

Moreover, the existing risk assessments related to APM often overlook the potential for the technology to counterbalance the very risks it might create. For example, while APM could theoretically enable a biohazard risk of highly enhanced pathogens, it would also offset this risk through the development of more effective cures, vaccines, and medical tools to combat such biohazards. However, for the other theoretical risks, their assumption-heavy nature fails to account for the physical constraints that may limit the feasibility of these extreme scenarios in the near future. Right now, a huge technological barrier to APM development is scaling up, where going from micro to macro in a bottom-up fashion leads to significantly more potential failing points (\citetb{APMImpact}).

Furthermore, the problem of missed opportunities when it comes to intermediate technologies is even more evident when considering the Collingridge Dilemma. At the early stages of development, there is a vast lack of information as to the field’s future, and the way to combat it with as little speculation as possible is only by extrapolating on current trends. On the contrary, by considering extreme, long-term futures we only enhance this uncertainty. Technologies such as catalysis and DNA origami get little mention in literature aimed at the future vision of APM but are, in reality, precisely what will enable this vision. Another aspect of the nature of the technologies that employ APM is that they are interdisciplinary, with applications and advancements in quantum computing, medicine, AI, and many other fields. The current literature does not yet specifically identify the different current applications which are crucial to predicting the future trajectory of APM and outlining where progress can and is being made. 

This paper attempts to address some of the problems mentioned above, which are an overemphasis on speculative scenarios in APM literature, neglect of intermediate technologies that drive the future of development, disconnect between speculative theoretical APM and the current state of technologies, lack of consideration of the new technologies that can mitigate future risks due to APM, assumptions and unrealistic expectations in literature advocating for APM or its risks, failure to address and compile APM risks and benefits due to its interdisciplinary nature, and a poor understanding of the future of the field which is necessary for regulation and future direction. 

Solving the problem of overemphasis of speculative scenarios in APM literature would allow for a more realistic and grounded discourse on APM, which would empower more researchers to focus on achievable, near-term goals rather than only having a distant, speculative vision. This would also improve regulation capabilities in the future. We achieve these goals by writing a comprehensive review of the current state of specific technologies related to APM and examining the current barriers to development. Furthermore, we extrapolate on the current developments in the field to provide a highly non-speculative, comprehensive analysis of future prospects for this technology.

This paper also reviews the current state of APM in different fields to address the disconnect between speculative theoretical APM and the current state of technologies, fostering collaboration between theorists and practitioners, and leading to the development of realistic plans for the direction of APM development. 

\section{The current state of APM}

We believe that to analyze future prospects with any precision we must first examine the work already being done in the field. The following is an analysis of the current state of progress on APM, split into categories by purpose and applications.

\subsection{DNA Origami for Medicine and Biosensing}

In living cells, processes create biological materials to near atomic precision - unlike manmade 3D structures that tend to lack such control. DNA origami is a significant application of APM based on self-assembly to create nanostructures. DNA origami is a bottom up manufacturing method that uses single-stranded DNA that folds into specific shapes through complementary base pairing. By designing the sequence of the DNA, it can be programmed to fold into a complex 2D or 3D structure with atomic precision(\citetb{rothemund2006folding}). 

Current developments in DNA origami have reached a stage where it is possible to create stable and highly intricate structures. Different geometries in any dimensions can be made to spec, and designs such as nanoscale boxes, gears, and even DNA rotors have now been created(\citetb{kosuri2019rotation}). Various types of DNA nanostructures, including double-crossover DNA tiles, $4 \times 4$ tiles, and three-point-star tiles have been fabricated to be assembled into higher order nanostructures such as nanotubes, nearly arbitrarily complex 3D patterns, and 2D lattices (\citetb{DNAorigami}). Furthermore, current DNA origami allows massively parallel synthesis of well-defined nanostructures (a picomole-scale synthesis generates $10^{12}$ copies of a product).

What started out as DNA origami in the form of folding a $\sim$7000 nucleotide long single-stranded DNA has since become a powerful tool that allows for hugely more versatile development of non-periodic structures and 3D structures with user defined symmetries. 

DNA origami has applications in medicine, where DNA origami nanostructures can be used as carriers for drugs, capable of transporting agents to targeted cells. This reduces side effects and increases treatment efficacy, specifically in cancer therapy. However, current research is making progress in strengthening nanostructures to be durable enough to survive in the human body. One area of progress in this is structural enhancements of these APM structures. Typical planar DNA origami nanostructures have around 200 uniquely addressable points in an area of 8000-10000 square nanometers. This allows for places that can attach different dyes, metals, and enzymes with atomic precision (\citetb{DNAmoeities}). Certain developments in functional attachments are currently in coatings and functional components. Metal, silica, lipid, or polymer coatings can be applied in DNA origami to enhance durability and strength, as well as being able to change certain properties of the material such as increasing electrical conductivity or biocompatibility. 

A comprehensive review of DNA origami (\citetb{DNAorigami}) explains that DNA structures can already also be engineered to be dynamic, responding to external stimuli. This is made possible from strand displacement reactions or conformational switches, which are molecular mechanisms that enable a DNA origami to change its shape and can react to the environment. Strand displacement reactions are a common mechanism for DNA origami. One DNA strand is replaced or displaced by another, causing a change in the structure’s overall conformation. An example of this is introducing a complementary strand to cause a structure that's unfolded to fold or vice versa. Conformational switches are triggered by external stimuli such as pH, temperature, light, and presence of specific ions or molecules. This has applications in biosensing, where molecule detection can trigger conformational switches due to binding to DNA. Another example is targeted drug delivery, where a conformational switch can be designed to only release a drug when a specific condition such as a pH value is met.

\subsection{Electronics}

Currently, some of the biggest benefits and use cases of APM lie in the electronics industry. With scaled-down components, the world's computational power is expected to keep rising rapidly in the near future, giving birth to more and more complex algorithms and technologies.

\subsubsection{Atomically Precise Circuitry}

Scanning Tunneling Microscopy and Atomic Force Microscopy have given rise to the ability to create atomically precise nanoelectronics, such as graphene nanoribbon transistors or silicon-based electronics and nanoscale semiconductors. According to (\citetb{siliconelectronics}), these technologies allow us to manipulate silicon surfaces at the atomic level to create electronics parts such as binary wires and logic gates. This is done through performing APM on hydrogen-terminated silicon surfaces, which are silicon surfaces that have hydrogen atoms bonded to them. This technology has been demonstrated to work to create new electronics that are much smaller, in addition to being inherently faster and more energy-efficient. This has opened the door to a new realm of miniaturization, as well as extending Moore's Law for a few more years.

It's not just the field of silicon electronics that is seeing promise. APM has enabled the creation of even smaller (on the order of 10nm) and more effective transistors using technologies such as Atomic Layer Deposition and Molecular Beam Epitaxy. These methods have led to high carrier mobility while overcoming issues like the lack of a bandgap and purity concerns. This has led to reduced power consumption, increased processing speed, and overall improved device performance (\citetb{graphene}). An example of this are graphene nanoribbon transistors that (when coated with Aluminum Oxide to prevent the degradation of contact resistance over multiple measurement cycles) have been shown to be effective for several thousands of full cycles. Another significant benefit to graphene nanoribbon transistors is the customizability they have when compared to traditional silicon electronics. For instance, graphene nanoribbon transistors can be made with atomic precision, allowing for precise control over their width, edge structure, and bandgap. This level and range of control allows for the design of specific electronic properties tailored for various applications such as high-speed electronic devices, flexible electronics, and advanced sensors. By tuning the bandgap and edge configuration of graphene nanoribbons, researchers can optimize these transistors for specific functions. For instance, in high-speed electronics, narrow graphene nanoribbons with precise edge terminations can enhance carrier mobility, resulting in faster speeds. The ability to control the material’s properties at the atomic level in flexible electronics guarantees performance even when the device is bent or stretched. Additionally, in advanced sensors, the precise control over electronic properties enables the creation of highly sensitive detectors for various chemical or biological substances. This atomic-level customization of graphene nanoribbons thus opens up new possibilities for creating cutting-edge technologies with tailored functionalities.

\subsubsection{Quantum Computing}\label{CurrentQuantumComputing}

There has also been recent research about the applicability of atomic-precision advanced manufacturing for Silicon based quantum computing (\citetb{quantumsiliconAPM}). The authors describe the benefits of using such an approach in the manufacturing of quantum computing schemes:
\\
\begin{adjustwidth}{1.5cm}{1.5cm}
\textit{"A materials synthesis method that we call atomic-precision advanced manufacturing (APAM), which is the only known route to tailor silicon nanoelectronics with full 3D atomic precision, is making an impact as a powerful prototyping tool for quantum computing. Quantum computing schemes using atomic ( $ ^{31}P $) spin qubits are compelling for future scale-up owing to long dephasing times, one- and two-qubit gates nearing high-fidelity thresholds for fault-tolerant quantum error correction, and emerging routes to manufacturing via proven Si foundry techniques. Multiqubit devices are challenging to fabricate by conventional means owing to tight interqubit pitches forced by short-range spin interactions, and APAM offers the required (Å-scale) precision to systematically investigate solutions."}\\
\end{adjustwidth}

The paper highlights the significant influence that APM is already having and could continue to have on quantum computing. 

Historically, a large issue with quantum computing were the intense heat requirements, needing vast machines just to cool miniscule computers. However, with recent advances, nanoscale cooling devices are becoming possible, allowing for smaller and thus more mainstream quantum computers (\citetb{APMForQuantumCooling}).

\subsubsection{Artificial Intelligence}\label{CurrentAI}

With recent advances in Deep Learning and Artificial Neural Networks, many applications for AI in APM have already been proposed (\citetb{AIAPM}). It has been shown that AI is set to play a crucial part in further development and use of APM through interpretation of data, nanoscale simulations, materials analysis and design of APM systems.

Using the developments outlined in \ref{CurrentQuantumComputing}, novel Quantum Neural Network architectures have been proposed to aid in further AI progress (\citetb{QCforAI}, \citetb{QuantumAIArchitectures}). The aforementioned cooling devices (\citetb{APMForQuantumCooling}) may also aid in cooling traditional-architecture computing centers, enabling further AI progress.

The rapid advances in micro- and nanoscale sensors due to development of increasingly capable APM (\citetb{nanosensors}) could provide modern-day AI systems with more precise real-world data, enhancing their capabilities in fields such as robotics and self-driving vehicles.

\subsection{Environment}\label{NCCurrent}

Atomically precise metal Nanoclusters (NCs) have unique properties such as high catalytic activity, selectivity, and well-defined structure, which make them very promising for developing catalysts that could have a large impact on the environment through intermediate technologies. This area of study received significantly more attention when gold, for a long time considered without catalytic activity, exhibited high catalytic activity as nanogold catalysts, sparking a "gold rush" in academia (\citetb{nanogold}). Since then, the exploration and development of NC catalysts have led to many promises in energy storage and generation technologies. 

First of all, there are many methods of creating atomically precise NCs, one of which is direct synthesis. By carefully controlling the reaction conditions such as reaction time and temperature, we can produce NCs with a specific number of atoms, thereby achieving atomic precision (\citetb{directsynth}). This is important because modern NCs have provided us with new technical capabilities in electrocatalysis, photocatalysis, and organic reactions. This paper only covers benefits in electrocatalysis, due to the highly technical nature of the other two.

Electrocatalysis involves accelerating electrochemical reactions that are crucial in energy conversion technologies such as fuel cells, batteries, and electrolysis. NCs are specifically powerful in their large surface areas and active sites. For example, NCs were shown to exhibit ultrahigh catalytic activity in fuel cells due to their huge specific surface areas and unique electronic structures (\citetb{electrocat}). Furthermore, in direct methanol fuel cells, the oxidation of methanol at the anode generates electricity. NCs, particularly those of platinum or alloyed with other metals, can catalyze this reaction more efficiently than bulk materials (\citetb{catalyzer}).

\section{The future of APM in areas of current development}

By extrapolating on the trends seen in the developments outlined above, it is possible to estimate the future of APM with reasonably high certainty. The following is our prediction for the risks and benefits that could arise from each venue of progress in the next few decades.

\subsection{Medicine and Biosensing}

The future of DNA origami in drug delivery is promising. DNA origami is revolutionary in its ability to have highly targeted and specific delivery systems due to its ability to be designed to respond to external stimuli. Combined with enhancements to DNA nanostructures such as molecules that bind only to certain cells, we can envision highly targeted medicine delivery systems. This could dramatically improve the effectiveness of treatments while reducing side effects. Additionally, the development of smart drug delivery systems that respond to environmental stimuli (such as changes in pH or the presence of specific biomarkers) will allow for a system that heals on command, and could remain passive until then.

Despite these advancements pointing towards promising future capabilities, current progress in DNA origami for drug delivery faces several barriers, although mostly technical. These challenges include integrating enhancements such as metals or enzymes to the structure as well as making very complex ones. Researchers are working on improving the stability of DNA origami structures and addressing challenges such as size limitations and scale of production. Progress in this field includes what was mentioned earlier about structural enhancements, which allows for the attachment of different dyes, metals, and enzymes to enhance strength or add certain properties. There are few physical barriers that we can see so far.

\subsection{Electronics}

\subsubsection{Quantum Computing}

APM has the potential to bring significant real-world benefits to quantum computing by addressing some of the most critical challenges faced by this emerging field.

One example of this is that APM could create materials with near-perfect atomic structures, reducing defects that can interfere with qubit coherence which severely limits the performance of quantum computers. This would ensure that each qubit is fabricated with atomic precision, leading to uniformity across qubits, reducing variability, and improving reliability.

APM could also allow for the creation of quantum components with minimal energy loss by optimizing atomic arrangements. This can lead to quantum systems that are more energy-efficient, which is particularly important as quantum computers scale up in size and complexity.

\subsubsection{Artificial Intelligence}

Much focus has been directed to AI development, a stark contrast to the minimal focus on methods for safe development of AI. An even more neglected area has been the adjacent technologies that could further facilitate AI development, which would certainly also have to be explored to decrease the existential risk of AI. We would argue that APM is an example of such a technology.

The creation of artificial general intelligence (AGI) or artificial super intelligence (ASI) seems to be the “holy grail” of AI research and computer science. Modern-day large language models (LLMs) have become popular in the previous years, further increasing the rate of development of the field. This has been shown through the increasing training computation needed to train those models. However, the computational ability of the top 500 supercomputers has not caught up with the size of modern day AIs, which could be one of the bottlenecks to the creation of more powerful LLMs or AGI/ASI. One of the potential uses of APM would be the manufacturing of smaller transistors with a low error rate at cheap costs and high numbers. Combined with the more energy efficient design of computers allowed by APM, it would create the perfect scenario for AI development. This idea has been backed up by the aforementioned research on the way APM could greatly benefit quantum computing and the application of quantum computers to AI (\citetb{QCforAI}).

\begin{figure}[H]
    \includegraphics[width=1\textwidth]{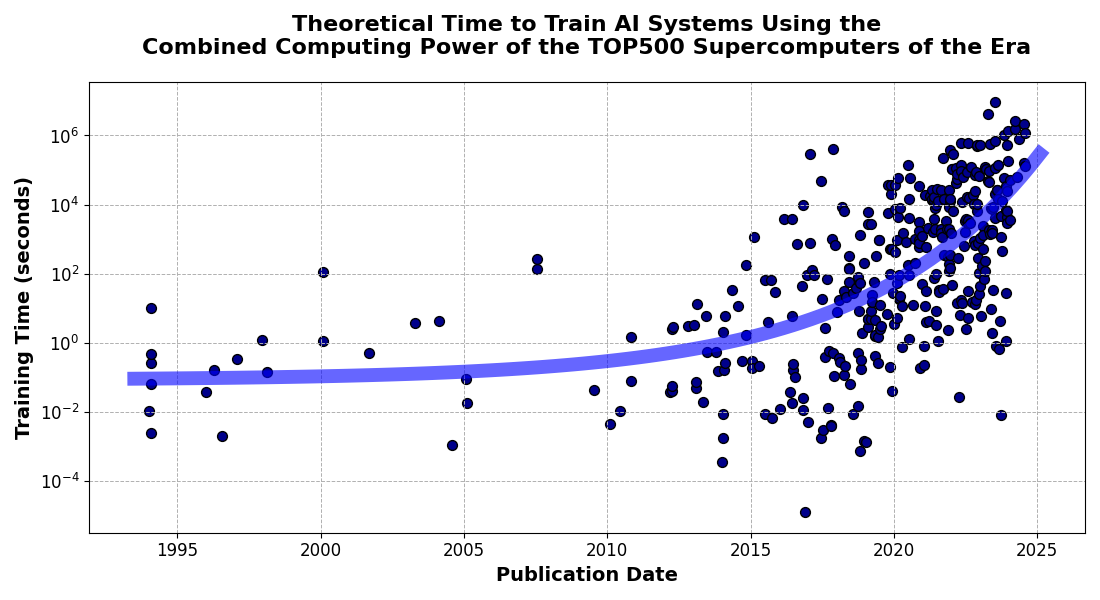}
    \caption{Data from (\citetb{AI}) and (\citetb{Supercomputer})}
    \label{AI}
\end{figure}

So, given a scenario where APM is developed before AGI/ASI, we argue it would aid the development of such technology. In order to analyze the impact of APM in relation to AI development it is necessary to assess the impact of AI itself, which is done extensively in other resources and also is not the focus of this paper. However, in the scenario where AGI/ASI is developed first, the impact of APM could be varied. Some argue that AI could decrease the risks associated with APM through the potential ability to manage, coordinate and decide using ethical frameworks in accordance with its training. Others argue that APM could be a tool used by malicious AI systems to further manipulate the environment and create harm. Another scenario could involve both technologies being developed at similar times and thus accelerating each other's development to form a positive feedback loop. We think this scenario is so likely and impactful that it deserves its own analysis and thus will not go into many details here.

In conclusion, estimating the net effect of APM in relation to AI has a lot of open subproblems, such as the net effect of AI itself. Therefore, only a preliminary conclusion can be made. If AI has a net positive effect on humanity, it is exceedingly likely that APM's impact will also be positive. Likewise, if AI has a net negative impact then APM would also have a net negative impact. This reasoning lies in the deep symbiotic connection formed between AI and APM, as described in \ref{CurrentAI}.

\subsection{Environment}

The potential of APM to transform environmental solutions is immense, particularly through advancements in nanocluster (NC) technologies. These advancements are paving the way for revolutionary changes in energy conversion technologies and energy generation, areas where NCs have already shown significant promise, as described in section \ref{NCCurrent}.

NCs are being explored as catalysts for converting carbon dioxide into useful chemicals, which would work to reduce carbon emissions. As APM technologies advance, we can expect NCs to become even more precise, further improving benefits mentioned in section \ref{NCCurrent}. This progress could lead to the development of new catalytic processes that are more environmentally friendly and cost-effective, further driving the transition to cleaner energy sources.

In terms of energy generation, NCs are already contributing to the enhancement of photovoltaic cells, improving light absorption and conversion efficiency (\citetb{catalyzer}). As APM advances, these improvements are likely to accelerate, which could lead to a new generation of solar cells, significantly reducing the global carbon footprint. Furthermore, future APM technologies could enable the production of  not only more efficient but also less environmentally harmful photovoltaics.

Furthermore, NC filters could be developed to capture and neutralize pollutants at the molecular level, addressing the problems of water and air contamination. However, APM also entails potential risks, such as nanopollution. As NC production increases, so does the potential for environmental contamination. However, current research is addressing these risks, with future APM technologies likely to include mechanisms for detecting and neutralizing nanoparticles as well as water and air contamination (\citetb{nanozyme}).

In conclusion, the future of APM holds great promise for advancing energy conversion and generation technologies, with NCs playing a pivotal role.

\section{The speculative future of APM}

We also recognize that some progress is likely to be made in subfields that emerge along the way, ones where little has been done so far. We are aware that this analysis is based much more on speculation than the previous, and we are much more uncertain about it. However, simply disregarding these areas of development would be quite imprudent due to their major theoretical impact, hence we’ve decided to still address them. Much of this analysis is inspired by (\citetb{Umbrello}), detailing their risk/benefit analysis for APM and expanding on their work for other areas.

\subsection{Material Wealth}

If APM is developed, one of the most clear ways it may be beneficial is through the cheap manufacturing of a large variety of items. Dr. Eric Drexler calls this “Radical Abundance”. He describes this as a utopian scenario in which food shortages are a thing of the past and even complex consumer items are widely distributed due to the miniscule manufacturing costs. We will expand on the benefits and downsides of such a scenario.

Dr. Eric Drexler, a researcher of nanotech, has expanded on this concept in his book: “Radical Abundance: How a Revolution in Nanotechnology Will Change Civilization” (\citetb{Drexler}). However, there is research in economics on the concept of diminishing marginal utility of consumption, the principle that a dollar is worth more to a poor person than a rich one. It is expected that richer countries will invest more into nanotech research, thus they will be leaders in the development/use of the technology. This could mean only the richer countries have access to the technology and poorer countries become dependent (at least initially). In the worst case scenario, APM could impoverish countries that have no place in the APM economy. This would be a particularly sharp example of technological unemployment, the idea that developments in technology and working practices cause some workers to lose their jobs.

Furthermore, like with AI, the impact of APM on the job market would be revolutionary. Many jobs in manufacturing and other related sectors could become useless in the advent of APM. There has also been debate over the role of material wealth on the quality of life of individuals with some research suggesting that wealth does not impact quality of life after some threshold (\citetb{MatWealth}), though this is debated. However, even if all these considerations are real possibilities it does not change the fact that as of 2022, almost 23\% of the global population (1.8 billion people) were living on less than \$3.65 a day and they would greatly benefit from a technology that expands their material wealth.

However, it is also not wise to assume that if APM is possible, all of its described benefits such as curing all diseases and drastic manufacturing cost reductions would happen. Factors such as the energy cost of performing APM could largely counteract the benefits to material wealth, and there may be complications that prevent curing certain diseases and cannot be foreseen at this time. 

In conclusion, APM has the potential to be of great benefit to quality of life, especially for people living in lower income countries. It also has the danger of being a tool used to further wealth inequality if misused. So, it is of utmost importance that if APM is pursued, it must be done so jointly by both high income and low income countries.

\subsection{Space Travel}

Though this factor is not of immediate importance for achieving humanity's long term goals, it is still relevant to discuss the large potential impact of APM on space travel. With the growing number of object launched into space (\citetb{Satelites}) and growing interest in developing permanent settlements on the Moon and Mars, there is little doubt that humanity's future resides outside the Earth.

However, the cost of space travel is fundamentally limited by the mass of components of the payload. With the development of APM, it could bring about massive innovation on lightweight materials and manufacturing systems and be of gigantic importance to the further development of the field of rocketry.

Another one of the direct ways APM could benefit space travel would be the facilitation of asteroid mining. There has been serious research on the feasibility of such an idea and if it would be economically viable with today's technology. The consensus is that while it is feasible to capture asteroid surface samples and return them to earth, as demonstrated by the OSIRIS-REx mission (\citetb{Asteroid}), it is much less viable than traditional mining techniques. This could change with APM as assembly systems would be miniaturized and ore treatment facilities could be manufactured on the spot. This could further facilitate the development of humanity's presence in space by being able to acquire materials in nearby locations.

Another, more speculative benefit of APM is facilitating large scale space energy production such as space-based solar power. This has been considered a real possibility by some governments (\citetb{NASA}) and could be achieved if APM brings a cheaper cost of space travel and also miniaturizes the manufacturing systems. This is argued to be plausible in principle. The development of advanced space-based energy production technologies could also make it easier to establish colonies on other celestial bodies, increasing the civilizational resilience of humanity.

In conclusion, the positives of APM in relation to space travel seem to be very high and the downsides appear to be extremely small, so (apart from the less pressing nature of advanced space travel) it is an excellent example of the positive impact APM could bring if properly developed.

\subsection{Mass Surveillance}

The concept of mass surveillance and its relation to security and personal privacy is a complex and controversial issue. This could be further complicated with the development of APM due to its capabilities to enhance/develop today's networks. We argue that APM could facilitate privacy-related s-risks via the development of mass surveillance.

Not only could APM enable the creation of new surveillance technologies (such as new sensors) but it could also increase the data processing capabilities of surveillance networks with new computers and AI systems. As with most APM-created items, the new visual/audio sensors would be very cheap to produce and extremely difficult to detect due to the small size. On one hand, facilitation or enhancement of mass surveillance  would be beneficial for decreasing crime and deterring other activities such as bioterrorism or illegal nuclear weapons development, decreasing x-risks. On the other hand, APM could also increase s-risks with a worst case scenario of an Orwellian totalitarian regime that uses mass surveillance to suppress any revolutions. 

The net effect of APM in this area would depend heavily on the countries in which it is developed and used. In places with regulations that tackle individual privacy, the effect of APM thereon could be minimal if those regulations are properly enforced. On the other hand, countries with vast surveillance networks would have no problem exacerbating APM’s effect. There are also the further tensions that could develop between superpowers as espionage becomes easily available and virtually undetectable.

In conclusion, APM can change the face of mass surveillance and introduce tension between nations due to easier espionage, while also being able to decrease x-risks due to the deterrence imposed by an “all seeing eye”. There is also disagreement on whether personal privacy should be sacrificed to decrease crime and x-risks. So overall, the net value cannot be calculated before describing the net value of mass surveillance itself.

\subsection{Gray Goo}

Another scenario worth considering is the notorious "gray goo" catastrophe where self-replicating nanobots exponentially grow in numbers, eventually consuming all the bio-matter on Earth. This has been reviewed countless times in  literature (\citetb{Graygoo}) relating to APM, and it has been deemed unlikely to occur by accident since the accidental design of a self-replicating nanobot that uses the right types of elements found in the earth's biosphere would be vanishingly small. This implies that gray goo would have to be of malicious intent. Furthermore, APM systems could be designed to not produce self-replicating systems or have background checks to see if the object being produced is some sort of self-replicator. Thus, it seems to be a very speculative and improbable issue overall.

\section{Conclusion}

In the rapidly progressing field of Atomically Precise Manufacturing (APM), a large amount of previous research has focused on speculative theoretical worst-case scenarios. Often outdated and plagued by assumptions, these papers are of limited value for further work at the current stage of development. To combat this problem, we have assembled a comprehensive review of the current state of APM. Through extrapolation of current trends accounting for the barriers to entry, we have put together a highly non-speculative vision of the near-term future of this technology. With applications in areas such as DNA Origami, Quantum Computing, Nanoscale Electronics and many more, it is evident that APM is becoming a critical technology with far-reaching prospects. Using these predictions as a basis, it also becomes much easier to speculate about the longer-term future of this technology in areas such as Mass Surveillance or Space Travel, and even consider more far-fetched possibilities such as an explosion of Material Wealth or a Gray Goo scenario.

With our analysis completed, we would like to see further, more in-depth research done on field-specific applications of APM and the potential they could have on humanity’s future. In the following years, once more advanced versions of APM are developed, governments should begin addressing these issues directly by using expert input to devise and refine policies, creating a foundation for years to come. By the end of the next decade, the policies should start being implemented on a global scale to ensure a positive trajectory for APM and related technologies.

\section{Acknowledgements}

This paper was written as part of the Non-Trivial Fellowship, and we are greatful to everyone that has helped us along our path. In particular, we thank Dr. Eric Drexler for an immensely useful conversation that has led us to pivot the direction of our project for the better.

\section{References}
\renewcommand\refname{}
\bibliographystyle{plainnat}
\bibliography{output}

\begin{thebibliography}{31}
\providecommand{\natexlab}[1]{#1}
\providecommand{\url}[1]{\texttt{#1}}
\expandafter\ifx\csname urlstyle\endcsname\relax
  \providecommand{\doi}[1]{doi: #1}\else
  \providecommand{\doi}{doi: \begingroup \urlstyle{rm}\Url}\fi

\bibitem[Abdelgaber and Nikolopoulos(2020)]{QCforAI}
Nahed Abdelgaber and Chris Nikolopoulos.
\newblock Overview on quantum computing and its applications in artificial intelligence.
\newblock In \emph{2020 IEEE Third International Conference on Artificial Intelligence and Knowledge Engineering (AIKE)}, pages 198--199, 2020.
\newblock \doi{10.1109/AIKE48582.2020.00038}.

\bibitem[Bussmann et~al.(2021)Bussmann, Butera, Owen, Randall, Rinaldi, Baczewski, and Misra]{quantumsiliconAPM}
Ezra Bussmann, Robert~E. Butera, James H.~G. Owen, John~N. Randall, Steven~M. Rinaldi, Andrew~D. Baczewski, and Shashank Misra.
\newblock Atomic-precision advanced manufacturing for si quantum computing.
\newblock \emph{MRS Bulletin}, 46:\penalty0 607--615, 2021.
\newblock URL \url{https://doi.org/10.1557/s43577-021-00139-8}.

\bibitem[Chang et~al.(2020)Chang, Morgan, Bedier, Chieng, G{\'o}mez, Raminani, and Wang]{nanosensors}
Megan Chang, Georgia Morgan, Fatima Bedier, Andy Chieng, Pedro Roberto~Rodr{\'i}guez G{\'o}mez, Sathya Raminani, and Yixian Wang.
\newblock Review—recent advances in nanosensors built with pre-pulled glass nanopipettes and their applications in chemical and biological sensing.
\newblock \emph{Journal of the Electrochemical Society}, 167, 2020.
\newblock URL \url{https://api.semanticscholar.org/CorpusID:213222663}.

\bibitem[Choi et~al.(2020)Choi, Oh, and Kim]{QuantumAIArchitectures}
Jaeho Choi, Seunghyeok Oh, and Joongheon Kim.
\newblock The useful quantum computing techniques for artificial intelligence engineers.
\newblock In \emph{2020 International Conference on Information Networking (ICOIN)}, pages 1--3, 2020.
\newblock \doi{10.1109/ICOIN48656.2020.9016555}.

\bibitem[Dey(2021)]{DNAorigami}
Swarup Dey.
\newblock Dna origami.
\newblock \emph{Nature Reviews Methods Primers}, 1, 2021.
\newblock URL \url{https://www.nature.com/articles/s43586-020-00009-8}.

\bibitem[Dinh(2024)]{graphene}
Christina Dinh.
\newblock Atomically precise graphene nanoribbon transistors with long-term stability and reliability.
\newblock \emph{ACS nano ASAP}, 2024.

\bibitem[Drexler(2013)]{Drexler}
K~Eric Drexler.
\newblock \emph{Radical abundance: How a revolution in nanotechnology will change civilization}.
\newblock Public Affairs, 2013.

\bibitem[Drexler(2015)]{drexlerconvo}
K~Eric Drexler.
\newblock A conversation with dr. eric drexler.
\newblock 2015.

\bibitem[Du et~al.(2020)Du, Sheng, Astruc, and Zhu]{catalyzer}
Yuanxin Du, Hongting Sheng, Didier Astruc, and Manzhou Zhu.
\newblock Atomically precise noble metal nanoclusters as efficient catalysts: A bridge between structure and properties.
\newblock \emph{Chemical Reviews}, 2020.

\bibitem[Elkomy et~al.(2024)Elkomy, El-Naggar, Elantary, Gamea, Ragab, Basyouni, Mouhamed, and Elnajjar]{nanozyme}
Hager~A. Elkomy, Shimaa~A. El-Naggar, Mariam~A. Elantary, Sherif~M. Gamea, Mahmoud~A. Ragab, Omar~M. Basyouni, Moustafa~S. Mouhamed, and Fares~F. Elnajjar.
\newblock Nanozyme as detector and remediator to environmental pollutants: between current situation and future prospective.
\newblock \emph{Environmental Science and Pollution Research}, 31\penalty0 (3):\penalty0 3435--3465, 2024.
\newblock ISSN 1614-7499.
\newblock \doi{10.1007/s11356-023-31429-0}.
\newblock URL \url{https://doi.org/10.1007/s11356-023-31429-0}.

\bibitem[Giattino et~al.(2023)Giattino, Mathieu, Samborska, and Roser]{AI}
Charlie Giattino, Edouard Mathieu, Veronika Samborska, and Max Roser.
\newblock Artificial intelligence.
\newblock \emph{Our World in Data}, 2023.
\newblock URL \url{https://ourworldindata.org/artificial-intelligence}.

\bibitem[Haruta et~al.(2006)Haruta, Kobayashi, Sano, and Yamada]{nanogold}
Masatake Haruta, Tetsuhiko Kobayashi, Hiroshi Sano, and Nobumasa Yamada.
\newblock {Novel Gold Catalysts for the Oxidation of Carbon Monoxide at a Temperature far Below 0 °C}.
\newblock \emph{Chemistry Letters}, 16\penalty0 (2):\penalty0 405--408, 03 2006.
\newblock ISSN 0366-7022.
\newblock \doi{10.1246/cl.1987.405}.
\newblock URL \url{https://doi.org/10.1246/cl.1987.405}.

\bibitem[Jacobstein(2010)]{Jacobstein2010ForesightGF}
Neil Jacobstein.
\newblock Foresight guidelines for responsible nanotechnology development.
\newblock 2010.
\newblock URL \url{https://api.semanticscholar.org/CorpusID:226253545}.

\bibitem[Kang et~al.(2020)Kang, Li, Zhu, and Jin]{directsynth}
Xi~Kang, Yingwei Li, Manzhou Zhu, and Rongchao Jin.
\newblock Atomically precise alloy nanoclusters: syntheses{,} structures{,} and properties.
\newblock \emph{Chem. Soc. Rev.}, 49:\penalty0 6443--6514, 2020.
\newblock \doi{10.1039/C9CS00633H}.
\newblock URL \url{http://dx.doi.org/10.1039/C9CS00633H}.

\bibitem[Kasser(2002)]{MatWealth}
Tim Kasser.
\newblock The high price of materialism, 2002.
\newblock URL \url{https://api.semanticscholar.org/CorpusID:145304136}.

\bibitem[Kosuri et~al.(2019)Kosuri, Altheimer, Dai, Yin, and Zhuang]{kosuri2019rotation}
Pallav Kosuri, Benjamin~D Altheimer, Mingjie Dai, Peng Yin, and Xiaowei Zhuang.
\newblock Rotation tracking of genome-processing enzymes using dna origami rotors.
\newblock \emph{Nature}, 572\penalty0 (7767):\penalty0 136--140, 2019.

\bibitem[Lauretta et~al.(2024)]{Asteroid}
Dante~S. Lauretta et~al.
\newblock Asteroid (101955) bennu in the laboratory: Properties of the sample collected by osiris-rex.
\newblock \emph{Meteoritics \& Planetary Science}, 2024.
\newblock URL \url{https://doi.org/10.1111/maps.14227}.

\bibitem[Liu et~al.(2017)Liu, Zheng, Du, Nasaruddin, Chen, and Xie]{electrocat}
Yanbiao Liu, Yuying Zheng, Bowen Du, Ricca~Rahman Nasaruddin, Tiankai Chen, and Jianping Xie.
\newblock Golden carbon nanotube membrane for continuous flow catalysis.
\newblock \emph{Industrial \& Engineering Chemistry Research}, 56\penalty0 (11):\penalty0 2999--3007, 2017.
\newblock \doi{10.1021/acs.iecr.7b00357}.
\newblock URL \url{https://doi.org/10.1021/acs.iecr.7b00357}.

\bibitem[Mathieu and Roser(2022)]{Satelites}
Edouard Mathieu and Max Roser.
\newblock Space exploration and satellites.
\newblock \emph{Our World in Data}, 2022.
\newblock URL \url{https://ourworldindata.org/space-exploration-satellites}.

\bibitem[Phoenix and Drexler(2004)]{Graygoo}
Chris Phoenix and Eric Drexler.
\newblock Safe exponential manufacturing.
\newblock \emph{Nanotechnology}, 15:\penalty0 869, 06 2004.
\newblock \doi{10.1088/0957-4484/15/8/001}.
\newblock URL \url{https://iopscience.iop.org/article/10.1088/0957-4484/15/8/001}.

\bibitem[Pitters et~al.(2024)Pitters, Croshaw, Achal, Livadaru, Ng, Lupoiu, Chutora, Huff, Walus, and Wolkow]{siliconelectronics}
Jason Pitters, Jeremiah Croshaw, Roshan Achal, Lucian Livadaru, Samuel Ng, Robert Lupoiu, Taras Chutora, Taleana Huff, Konrad Walus, and Robert~A. Wolkow.
\newblock Atomically precise manufacturing of silicon electronics.
\newblock \emph{ACS Nano}, 18\penalty0 (9):\penalty0 6766--6816, 2024.
\newblock \doi{10.1021/acsnano.3c10412}.
\newblock URL \url{https://doi.org/10.1021/acsnano.3c10412}.
\newblock PMID: 38376086.

\bibitem[Randall et~al.(2012)Randall, Ehr, Ballard, Owen, Saini, Fuchs, Xu, and Chen]{APMImpact}
John Randall, James Ehr, Joshua Ballard, James Owen, Rahul Saini, Ehud Fuchs, Hai Xu, and Shi Chen.
\newblock Atomically precise manufacturing: The opportunity, challenges, and impact.
\newblock \emph{Atomic Scale Interconnection Machines, Advances in Atom and Single Molecule Machines. ISBN 978-3-642-28171-6. Springer-Verlag Berlin Heidelberg, 2012, p. 89}, pages 89--, 04 2012.
\newblock \doi{10.1007/978-3-642-28172-3_7}.

\bibitem[Rodgers et~al.(2024)Rodgers, Gertsen, Sotudeh, Mullins, Hernandez, Le, Smith, and Joseph]{NASA}
Erica Rodgers, Ellen Gertsen, Jordan Sotudeh, Carie Mullins, Amanda Hernandez, Hanh~Nguyen Le, Phil Smith, and Nikolai Joseph.
\newblock Space-based solar power.
\newblock Technical report, NASA Office of Technology, Policy, and Strategy, 2024.
\newblock URL \url{https://ntrs.nasa.gov/citations/20230018600}.

\bibitem[Rothemund(2006)]{rothemund2006folding}
Paul~WK Rothemund.
\newblock Folding dna to create nanoscale shapes and patterns.
\newblock \emph{Nature}, 440\penalty0 (7082):\penalty0 297--302, 2006.

\bibitem[Sacha and Varona(2013)]{AIAPM}
G.M Sacha and P.~Varona.
\newblock Articicial intelligence in nanotechnology, 2013.
\newblock URL \url{https://repositorio.uam.es/bitstream/handle/10486/665596/artificial_sacha_NT_2013_ps.pdf}.

\bibitem[Seeman(2017)]{DNAmoeities}
Nadrian~C Seeman.
\newblock Dna nanotechnology.
\newblock \emph{Nature Reviews Materials 3.1}, 2017.
\newblock URL \url{https://www.nature.com/articles/natrevmats201768}.

\bibitem[Snodin(2022)]{APMclasses}
Ben Snodin.
\newblock Nanotechnology strategy research as an ea cause area, 2022.
\newblock URL \url{https://forum.effectivealtruism.org/posts/oqBJk2Ae3RBegtFfn/}.

\bibitem[Strohmaier et~al.(2024)]{Supercomputer}
Erich Strohmaier et~al.
\newblock Top500, 2024.
\newblock URL \url{https://top500.org/statistics/perfdevel/}.

\bibitem[Takhistov(2016)]{ProtonDecay}
Volodymyr Takhistov.
\newblock Review of nucleon decay searches at super-kamiokande, 2016.
\newblock URL \url{https://arxiv.org/abs/1605.03235}.

\bibitem[Tan et~al.(2016)Tan, Partanen, Lake, Govenius, Masuda, and M{\"o}tt{\"o}nen]{APMForQuantumCooling}
Kuan~Yen Tan, Matti Partanen, Russell~E. Lake, Joonas Govenius, Shumpei Masuda, and Mikko M{\"o}tt{\"o}nen.
\newblock Quantum-circuit refrigerator.
\newblock \emph{Nature Communications}, 8, 2016.
\newblock URL \url{https://api.semanticscholar.org/CorpusID:12688487}.

\bibitem[Umbrello and Baum(2018)]{Umbrello}
Steven Umbrello and Seth~D. Baum.
\newblock Evaluating future nanotechnology: The net societal impacts of atomically precise manufacturing, 2018.
\newblock ISSN 0016-3287.
\newblock URL \url{https://www.sciencedirect.com/science/article/pii/S0016328717301908}.

\end{thebibliography}

\newpage
\appendix
\pagenumbering{roman}

\section{Appendices}

\subsection{Quantum Effects}

It is common practice to disregard quantum-scale phenomena when dealing with atomic-scale objects due to their low frequency of occurrence and virtually negligible impact, but we consider it important to explain why this is the case, the reason it is not necessary to account for their impacts.

\subsubsection{Proton Decay}

For instance, consider proton decay. Since the lower bound for the lifetime of a proton is currently estimated to be on the order of $\tau\approx10^{33}$ years (Some research suggests that this estimate is too short (\citetb{ProtonDecay})), we can model the probability of observing a proton decay in any given time of observation by considering the decay equation:

\begin{equation} N = N_0 e ^ {( - \lambda t )} \end{equation}

Where $\lambda = 1 / \tau \approx 1 / 10 ^ {33}$ per year and $N_0$ is the number of protons being observed at the start, we expect $N$ to be the number of protons left after observing for a time of $t$. Since the decay constant ($\lambda$) is so small, the exponential $e^{(-\lambda t)}$ can be represented by the first two terms of the exponential series:
\\\\
$e ^ {( - \lambda t )} \approx 1 - \lambda t$\\
\\
$N \approx N_0 ( 1 - \lambda t )$\\
\\
$N_0 - N \approx N_0 - N_0 ( 1 - \lambda t ) \approx N_0 \lambda t$
\\\\
A highly advanced complex APM system that is capable of handling large amounts of matter extremely quickly (1kg of matter per 1kg of machinery per week) will only encounter at most $\frac{360}{7\times m_p} \approx 30.2 \times 10^{27}$  protons per year, where $m_p \approx 1.672 \times 10^{-33}$ kg is the mass of a single proton (This assumes that all matter handled is made purely of protons, which is impossible and is considered here only as a higher-edge limit). Assuming 100 years of non-stop work by 100 kg of machinery (AND all the upper-limit conditions from above), the expected number of observed proton decays can be represented by the following:
\\\\
$t = 100$\\
\\
$N_0 = 100 \times 30.2 \times 10^{27} = 30.2 \times 10^{29}$\\
\\
$N_0 - N \approx N_0 \lambda  t \approx \frac{30.2 \times 10^{29}}{10^{33}} \times 100 \approx 0.3$
\\\\
To rephrase this, the upper-limit probability of detecting ONE proton decay within 100 years of non-stop work by 100 kg of incredibly advanced machinery with matter made entirely of protons is still only about $30\%$ - understandably negligible.

\subsubsection{Heisenberg Uncertainty Principle}

Another often overlooked effect at this scale is the Heisenberg Uncertainty Principle. The equation reads:
\\\\
\begin{equation} \Delta(x) \Delta(p) \geq \frac{h}{4\pi} \end{equation}
\\\\
Where $\Delta(x)$ is the uncertainty in position, $\Delta(p)$ is the uncertainty in momentum and h is the Planck constant.
\\\\
In the case of APM, $\Delta(x)\approx 10^{-10}$ m and the mass of each single atom is $\sim 10^{-25}$ kg. Using the relation above, $\Delta(p) \geq 10^{-25}$ kgm/s. Thus, if (as a simplification) $p$ is taken to be the classical result $p = mv$, then $\Delta(v) \geq 1$ m/s which is negligible since the material would have an associated temperature of $\sim 0.001$K


\subsection{APM Classification (Legacy)}

\subsubsection{The GMCD}

During the first part of our research, we have come up with a novel idea for classifying APM into several categories based on the capabilities of such machinery. Similar ideas have been proposed before (\citetb{APMclasses}), but we found them to be exceedingly ambiguous and of limited value for future research. Thus, we have created the diagram you may see below to remove most ambiguity from these definitions. We planned to use the Classes we have defined for APM as guidelines for our risk analysis and base most of our further research around them.

The Generalized Manufacturing Class Diagram ($GMCD$) puts clear boundaries on classes of 3-dimensional high-precision manufacturing, near-APM and APM. Although there are many parameters that define an APM-capable machine (e.g. the price, the materials that can be handled, its availability to the masses, etc.), we think the two most important metrics are precision and throughput. These two measures define the capabilities of a machine as well as the risks posed by it to a high degree of certainty (Albeit the use cases may vary vastly depending on the other metrics).

We define precision as the smallest increment by which the intended position of a building block can be purposefully adjusted in all three dimensions using the native architecture of the manufacturing machine. In the $GMCD$, the red line signifies the boundary of APM and near-APM judging solely by precision. Since different use cases may see this to be anywhere between $3-4nm$ and $15-20nm$, we have selected this line to represent the median expectation of $10-12nm$.

As for throughput, the second metric used, we believe there are two definitions that may coexist and be used simultaneously. First, it is possible to calculate the throughput rate per mass of machinery ($R$) as $\frac{I}{TM}$, where $I$ is the total mass of inventory produced, $T$ is the time of production and $M$ is the total mass of machinery. This provides an extremely accurate measurement with very little calculation required. However, getting a precise result using this method will require a fully functional system doing work over an extended period of time, which we recognize to not always be an available possibility. As an alternative, we propose the following formula to calculate throughput for top-down high-precision machines:
\\\\
\begin{equation}
R = \frac{1-r}{M(t+rw)}
\end{equation}
\\\\ 
Where:\\\\
$R$ is the error-adjusted throughput rate per mass of machinery per unit of time\\
$r$ is the error rate of the machine, a value between 0 and 1\\
$t$ is the time the machine needs to place a single building block\\
$w$ is the time the machine needs to remove a misplaced block (fix an error)\\
$M$ is the mass of machinery\\

To derive this, let $E$ be the error-adjusted expected time needed to place a block. Then,
\\\\
$E = (1 - r)t + r(t + w + E) = t + rw +rE$
\\\\
Solving for $E$, we get:
\\\\
$E - rE = t + rw$\\
\\
$E (1-r) = t + rw$\\
\\
$E = \frac{t+rw}{1-r}$
\\\\
Since the total error-adjusted throughput rate per unit of time is 1/E, we can adjust for the mass of machinery and get:
\\\\
$R = \frac{1}{EM}$
\\\\
Which yields the aforementioned formula, 
\\\\
$R = \frac{1-r}{M(t+rw)}$
\\\\
Any combination of throughput and precision of a machine defines a point on the $GMCD$, sorting it into one of 6 classes (more about each of which can be found below). For now, the boundaries are very loose and merely guidelines, and we expect them to shift over time and become more and more solidified with new developments. We are also aware that the boundaries should be seen as a gradient, but for purposes of research and regulation it is beneficial to have them rigidly defined.

At a scale below $\sim0.1nm$, we expect atomic lattice limitations and quantum effects to render APM-capable devices useless, requiring an utterly different approach. This approach is virtually impossible to discuss at this point of development, and we do not expect a simple horizontal extrapolation of the diagram to a lower precision to be accurate. On the contrary, we do expect a vertical extrapolation of the $GMCD$ to account for throughput values that are not represented in the diagram to be reasonably precise.

\begin{figure}[H]
    \includegraphics[width=1\textwidth]{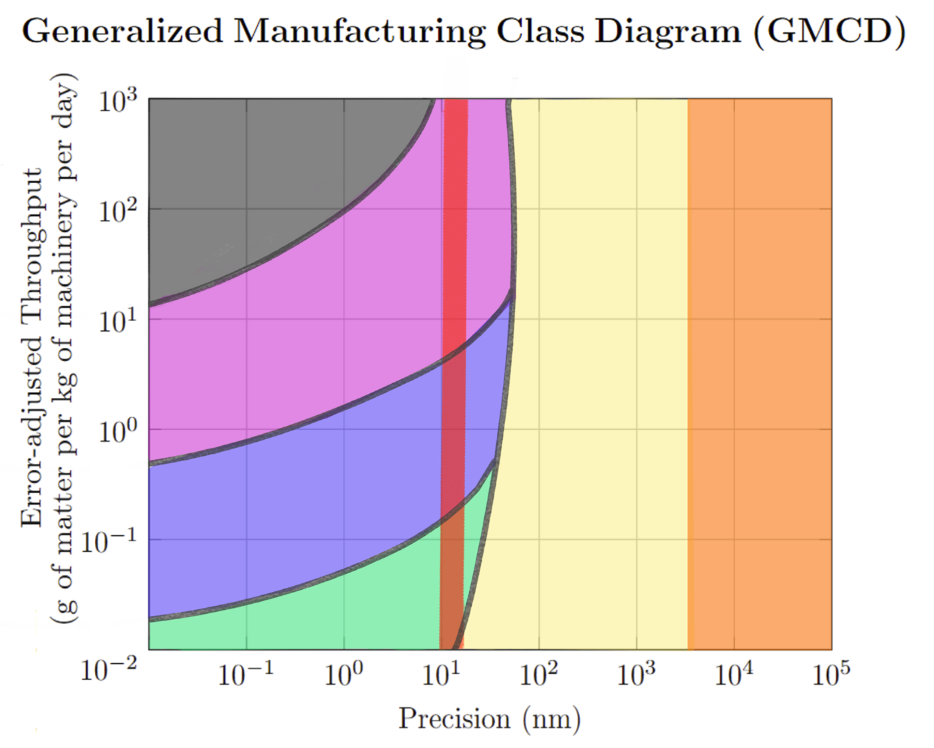}
    \caption{GMCD}
    \label{GMCD}
\end{figure}

\colorbox{YellowOrange}{Orange}: Normal manufacturing. This encompasses any semi- or fully automated manufacturing at a scale greater than $5 \mu m$. This includes the vast majority of all manufacturing occurring today and is heavily researched.

\colorbox{Goldenrod}{Yellow}: Precision machining. This encompasses any semi- or fully automated manufacturing with an error rate-adjusted precision of no more than $5 \mu m$. As of right now, such technology is fairly widespread, but the throughput of devices capable of such precision is relatively low. This area is also relatively popular with research.

\colorbox{YellowGreen}{Green}: Basic APM and Near-APM. This encompasses the early stages of APM-capable technology. Defined by atomic-scale precision and low throughput, the technology that falls into this category is likely to be incredibly expensive and only used for research. Today, the only instruments that could be argued to be in this class are scanning probe microscopes only available at an incredibly high price to a select few researchers. The risks posed by such technology would be very research-based; notable examples would include a gray goo scenario, novel virus manufacturing and early nanoscale weapons.

\colorbox{Cerulean}{Blue}: Technical APM and Near-APM. This class is characterized by atomic-scale precision as well as comparatively low throughput. We expect the use cases for such technology to be mostly research-driven due to the low throughput; however, it is likely that at this stage of development it will be available to the masses (albeit for a high cost). It is widely accepted within academia that such technology is possible, yet the timelines for its creation vary drastically. The risks from such technology would be more advanced versions of those from Basic APM and Near-APM as well as heavily toned-down versions of some of the risks from Mass and Complex APM.

\colorbox{Orchid}{Purple}: Mass APM and Near-APM. This category is defined by medium to high throughput and atomic-scale precision. Such technology is likely to drastically alter the world when (and if) it is achieved, with machines expected to be available to the public at a reasonable price. Most researchers agree that such technology is feasible in theory, but timelines are subject to major academic debate. The risks from such technology would be very similar to those from Complex APM, albeit be present at a scale that is less widespread and global but more narrow and local.

\colorbox{gray}{Black}: Complex APM. The feasibility of such technology is widely disputed, and it is common to assume that is is exceedingly unlikely to be achieved in the next few decades. Such machines are defined by extremely high throughput and low price (No less than $1kg$ of matter for every $1kg$ of machinery per day for no more than $\$1,000$) along with atomic and subatomic precision. The benefits and risks from Complex APM are what this paper is aimed at predicting, as well as steering future development to account for these risks and achieve a desirable future.

\subsubsection{Concerns}

Ultimately, after talking with Dr. Eric Drexler, we have realized that this diagram is of limited value. He has raised three major concerns about its utility.

The first concern is about the factors we used to classify APM-capable machines. It is quite possible to argue that precision and throughput are not, in fact, the main defining elements of such devices but that they should be characterized more abstractly by considering the items they are able to produce and the ways in which they influence matter. If this is true, it makes drawing unambiguous lines between categories virtually impossible. However, since there might be other ways to describe similar machinery, we could argue that the diagram may be remade in the future with different variables as the defining factors or even in more dimensions to account for this fallacy. On the basis of uncertainty, we have concluded this concern to be by far the lesser of the three, despite still being a major consideration.

The second concern lies in the unpredictable nature of the intermediate stages of the development of transformative technology. From past history, we see that it is much more difficult to make correct guesses about the early stages of a future technology’s development than about its final forms. Thus, it is possible that the classification of the intermediate stages of development that is presented in this diagram is invalid or worthless. However, tying back into our argument against the first concern, we expect the diagram to be altered and edited countless times in the future, allowing the change of numbers and even variables to account for current uncertainty. Therefore, we consider this argument to be slightly more significant than the first yet far less than the third.

The third and most significant concern lies in the “Snowball effect”, a positive feedback loop for technological coevolution. As described prior in this paper, we think APM has significant potential for helping humanity achieve AGI and ASI, as well as speeding up AI development in general. After a certain threshold, we believe AI will also contribute vastly to developing more sophisticated APM-capable machinery, thus creating a positive feedback loop: better APM creates better hardware and allows for training of better AI, while better AI can research and design better APM. This implies that once the threshold is reached, further development of both APM and AI will be sped up so significantly that all previous progress will round to zero within only a few years. Thus, even if the classification is completely correct, the intermediate stages that are classified in this diagram may only exist for a very brief period of time, making further analysis of them unimportant.

It was the last concern that has convinced us that this diagram is, exceedingly likely, not very valuable. However, we have spent quite a reasonable portion of time working on it and we do still think there could be use cases for the idea. So, we are still publishing it in case it does end up useful, however low the chance of that may be.

\end{document}